\documentclass[aps,prl,twocolumn,showpacs]{revtex4-1}
\usepackage{comment}
\usepackage{amsmath,amssymb,latexsym,mathrsfs}
\usepackage{graphicx} % Include figure files
\usepackage{color}
\usepackage{accents}
\usepackage{float}
\usepackage{gensymb}
\usepackage{tikz-feynman}
\begin{document}
\newcommand{\be}{\begin{eqnarray}}
\newcommand{\ee}{\end{eqnarray}}
\newcommand{\bs}{\boldsymbol}
\title
{\bf The anomalous spin texture as the probe for interactions in $Bi_{2-y} Sb_y Se_x Te_{3-x}$}
\author{T. Hakio\u{g}lu}
\affiliation{ Energy Institute and Department of Physics, 
{\.{I}stanbul Technical University, Maslak 34469, \.{I}stanbul, Turkey}
%\\
%and 
%\\
%Department of Physics, Boston University, Massachussetts, USA
}
\begin{abstract} 
The surface state of a three dimensional strong topological insulator (TI) is well described in the independent particle picture (IPP) by an isotropic Dirac cone at the $\Gamma$-point and perpendicular spin-momentum locking. Away from this point, the crystal point group symmetry causes anisotropic effects on the surface spectrum where a number of unusual effects are experimentally observed. In particular, the perturbative violations of the perpendicular spin-momentum locking frequently observed in many experiments remains to be a poorly understood feature theoretically. In parallel, the existence of electron-phonon interaction has been unquestionably verified by a number of experimental groups. In this article, we device an interacting theory of the spin texture using the spin-dependent self-energy formalism. We observe that the interactions lead to the observable spin-texture anomalies in the presence of a Fermi surface anisotropy while weakly affecting the energy bands. In particular, the experimental observation of the six-fold symmetric modulation of the in-plane spin in the $Bi_{2-y}Sb_ySe_{3-x}Te_x$ family and the resulting violation of the spin-momentum locking is explained using the coupling of an optical surface phonon to the surface electrons reported in earlier experiments. We also discuss recent puzzling results of the out-of-plane spin polarization experiments in this context. 

Our results introduce an interacting approach to the spin-related anomalies where the anisotropy of the Dirac cone around the Fermi surface proves to have an unconventional role. New experiments reporting unusual spin orientations in other materials with different symmetries signify that the theory introduced here may be relevant to a larger set of Dirac materials.
\end{abstract}
\maketitle

\section{introduction}
Three dimensional TIs with a single Dirac cone on their surface have taken great attention in the last 10 years due to their intriguing properties\cite{TI1,TI2,BiTe,HW_BiSeTe,BiSbTeSe}. Due to the strong spin-orbit coupling (SOC) the electronic band structure of a TI is inverted in the bulk. On the surface, the linearly dispersing states cross each other at the $\Gamma$ point forming a conducting spectrum and a spin-1/2 vortex with the topology of a $\pi$-Berry phase. The time reversal invariance (TRI) and the inverted bulk gap forbid the backscattering as well as the hybridization between the surface states leading to the protection of the topology\cite{backscattering}. These unique features make this family of TIs a promising research area not only in their potential device applications ranging from quantum computers\cite{qcomp} to new dissipationless electronic devices\cite{dissipationless_electronics}, but also in the exotic physics they capture\cite{TI1,TI2,exotic_physics}. 

From the theoretical point of view, the independent particle picture (IPP) has been overwhelmingly successful in understanding many of the properties of the TIs. In this picture, the surface states of a three-dimensional TI form a Dirac-like linear spectrum due to the strong SOC represented by an isotropic cone around the 
$\Gamma$ point. The spin is accurately defined by an in-plane component and is locked perpendicularly to the momentum defining a chiral spin-orbit state. As the electronic properties around this point are independent from the point group symmetry, much about new intriguing features of the surface state are found away from the $\Gamma$ point where the crystal symmetry becomes important. The first of such corrections within the single particle picture was discovered in the six-folded anisotropy of the Fermi surface in $Bi_2Te_3$\cite{BiTe} evolving from a perfect circle to a hexagonal shape and then into a snowflake pattern as the energy is increased away from the $\Gamma$ point. The similarly anisotropic Fermi surfaces were also observed in the whole class $Bi_2Se_xTe_{3-x}$\cite{HW_BiSeTe} and more recently in $Bi_{2-y}Sb_ySe_xTe_{3-x}$\cite{BiSbTeSe}. Remembering the six fold symmetry in the strongly spin-orbit coupled surface of Au-(111), a crucial difference is that, while in gold the six fold anisotropy is in all spin-independent and -dependent sectors, the strong warping of the Fermi surface in $Bi_2Te_3$ was discovered to have the origin in a highly nonlinear Dresselhaus type bulk SOC affecting the spin of the surface state as a result of which a significant out-of-plane spin component arises\cite{Fu_1}. 

It is known that Fermi surface anisotropy causes a number of genuine effects. Raghu and coworkers\cite{Raghu} showed that the electric transport and the spin dynamics is related in the vicinity of the $\Gamma$ point by a robust operator auto-correlation identity. Any anisotropic deviation from the continuous rotational symmetry, such as the hexagonal warping\cite{BiTe}, causes a violation of this operator identity. The HW combined with the scattering interactions can also introduce anisotropy in the scattering rates\cite{Pan,Valla, Barriga}. Another interest in anisotropy is driven by the motivation to engineer anisotropic Dirac cones to enable the carriers propagate with different Fermi velocities in different directions, creating an additional tunability for new applications. While various theoretical approaches have been proposed to make the anisotropic Dirac cones for the graphene, so far it has not met with success. There are also some theoretical predictions and/or experimental indications of anisotropy in other novel Dirac materials such as $AMnBi_2, (A = Sr, Ca)$ and $BaFe_2As_2$ but more experimental investigations are needed\cite{Ya_Feng,Richard}.   

The Dirac cone dynamics of the $Bi_2(Te_{3-x}Se_x)$ has been extensively studied as a function of $x$\cite{Chayou-Chen}. Since this series is a topological insulator for all $x\le 3$ with the same crystal structure, the systematic changes observed in the electron dynamics as a function of $x$ points at the phenomena that goes beyond the IPP. In this context, the electron-phonon interaction (EPI) coupling has been investigated\cite{Chayou-Chen}. The strength of the coupling depends on $x$, while the characteristic phonon frequency stays relatively unchanged. 

The IPP has been highly successful in describing the physical properties of the TIs and in many circumstances it is commonly agreed to ignore the interaction effects in the bulk and the surface bands. This is partly due to the strongly suppressed backscattering, an important element in the protection of the topological surface states by the TRI\cite{backscattering}. The absence of the backscattering however, does not exclude multiple finite angle scatterings which can be due to the impurities or some effective interaction mediated by collective excitations. In this context, coupling of the surface electrons to a number of excitations such as, spin density waves at the Fermi surface\cite{Fu_1}, spin plasmons\cite{Raghu,spin_plasmon} have been considered. The EPI has been heavily studied by a number of experimental\cite{Chayou-Chen,EXP_eph_TI,Kondo_Nakashima,Batanouny1,Batanouny2,Batanouny3} and theoretical\cite{TH_e-ph_TI,Thalmeier,Sarma} works where the Dirac quasiparticle lifetime, the self-energy corrections to the energy bands and the phonon frequencies have been estimated. The estimates of the coupling constants vary in a large range making difficult to extract the individual mode-specific phonons. Currently there is an increasing experimental evidence that the long-wavelength optical phonon modes in the sub $10 meV$ range play an important role in the EPI of the surface state\cite{Batanouny2,Batanouny3,Sobota}.    

Most of these experimental and theoretical works on interaction effects focused on the surface electron self-energy (SE). There, the finite quasiparticle lifetime and renormalization of the surface state energy were extracted from the complex SE corrections by measuring the surface electron momentum-distribution-curves (MDCs)\cite{Chayou-Chen,EXP_eph_TI,Kondo_Nakashima}. The complex SE provides broad information about the electron dynamics and is usually contributed by three main mechanisms: electron-electron, electron-phonon and electron-disorder interactions\cite{Chayou-Chen}. The electron-disorder interaction remains the dominant mechanism affecting the low energies and low temperatures where the e-ph interaction is suppressed. The coherent helium-atom-scattering experiments\cite{Batanouny2,Batanouny1,Batanouny3,Batanouny4} in a wide range of temperatures ($100 < T (K) < 300$) indicated the presence of the optical EPI pointing at a strong softening of the $7meV$ phonon mode. The position of the softening at the phonon momentum near $2k_F$ points at a Kohn anomaly and thus a strong coupling of the optical branch to the electronic states\cite{Batanouny4} with a less significant role played by the acoustic phonons\cite{Zhu,Pan,Heid}.   

In the presence of a strong SOC, the full information about the SE, is contained in a $2\times 2$ matrix, and in reality, the complex self energy extracted from the MDCs is a mixture of these components. The trace of this matrix comprises the spin-neutral part of the SE, whereas the rest of the components are combined in a covariant vector which rotates under rotations in the spin space. It was found that, the off-diagonal components of the SE play an important role in the vicinity of the $\Gamma$ point\cite{Aguilera}. The spin-dependent components are also expected to be important away from the $\Gamma$-point. This arises from the influence of the ion cores in the electronic subsystem providing a source for deviations from the isotropic Dirac-cone\cite{anisotropy}. The two fundamental outcomes are the deviations from the isotropic constant energy contours and an out-of-plane component of the spin. These two outcomes have both been observed in a number of Dirac materials irrespective of their topologies\cite{anisotropy}. It was also shown that, the emerging band anisotropy and the hybridization with the bulk bands in a spin-orbit coupled system have non-trivial effects on the in-plane spin-texture\cite{TH_STA}. It is therefore important to understand the individual role of each component in the SE in the spin dynamics. 

Looking at the spin-texture in the close vicinity of the $\Gamma$ point, the spin and the momentum are locked perpendicularly around the isotropic constant energy contours. This feature is common to all TIs and stands as one of the accurate predictions of the independent-particle k.p theory. Away from the $\Gamma$-point, violations of the perpendicular spin-momentum configuration has been recently observed in ARPES experiments for $Bi_2Se_3$ using circularly polarized light\cite{Y.H.Wang}. Similar violations were also reported in the ab-initio calculations for $Bi_2Te_3$\cite{Mirhosseini}. It was found that, other Dirac materials\cite{Bihlmayer}, in other symmetries\cite{Scholz}, and even in other materials such as quantum well heterostructures\cite{Bian} can also exhibit similar unusual features. In a broad perspective, it is therefore necessary to understand whether the effect of the interactions can extend beyond the quasiparticle energy spectrum, into the domain of the spin and the role played by the spin-dependent components of the SE.    

A clue was provided by the experiments in Ref.\cite{Y.H.Wang}, where several puzzling features of the spin were reported for the first time in the weakly warped Dirac cone of $Bi_2Se_3$. Writing the spin as ${\bs S}({\bs k})={\bs S}_{xy}+S_{z}\,\hat{\bs z}$, with $xy$ representing the topological surface plane and $\hat{\bs z}$ the normal to that plane, the $S_{z}$ is found to have a $2\pi/3$-periodic structure whereas the in-plane spin ${\bs S}_{xy}=S_{g}\,\hat{\bs g}_{\bs k}+S_{k}\,\hat{\bs k}$ has the regular perpendicularly spin-momentum locked component $S_{g}$ and additionally an anomalous component $S_{k}$ parallel to the momentum. The regular component is strong due to the strong SOC whereas the $S_{k}$ is weak and has a six-folded modulation disappearing along the high symmetry directions $\Gamma M$ and $\Gamma K$. A finite $S_{k}$ is unusual since it breaks the orthogonal locking between the spin and the momentum in striking contrast with the predictions of the k.p theory. Similar effects were observed also in the Rashba-split metals (eg. Bi/Ag (111))\cite{Meier}.    

Shortly after the publication of the Ref.\cite{Y.H.Wang}, the authors of Ref.\cite{Basak} proposed a model for this anomaly by adding to the warped Hamiltonian a new spin-orbit interaction fifth-order in momentum in the off-diagonal spin channel. 
Their basic  motivation is to reproduce the key features of the non-orthogonal state and its 6-fold symmetry by staying within the independent particle picture. It must be remembered that this family of materials has been known as good thermoelectric materials since 1950s\cite{Thermoelectric1,Thermoelectric2}. Interestingly, the presence of this term needs more experimental verification, since it has not been known not only after the more recent discovery of its topological properties but also before in its long history as a thermoelectric material. 

Other than the Ref.\cite{Basak}, the importance of this spin-texture anomaly (STA) in Ref.\cite{Y.H.Wang} was unnoticed until recently\cite{TH_STA}. There it is stressed that, the experimental observation of the STA is an important evidence for a need to go beyond the IPP. The existence of interactions was already known in these materials much earlier in their long history of thermoelectricity\cite{Thermoelectric2}. We consider these facts as a guiding motivation of this work in order to develop an interacting theory of the spin in these materials.

Confining our attention in this interacting formalism to the spin dependent sector of the SE, we show that, an anisotropy at the Fermi level is necessary to unlock the perpendicular spin-momentum configuration away from the $\Gamma$-point. The hexagonal warping (HW) observed in $Bi_2Se_xTe_{3-x}$ is known to bring anisotropy in the broadening of the MDCs with an asymmetry between the $\bar{\Gamma}\bar{M}$ and the $\bar{\Gamma}\bar{K}$ directions\cite{anisotropy}. It was previously proposed that\cite{TH_STA}, the interactions, the strong SOC and the hexagonal Fermi-surface anisotropy cooperate in a new unconventional mechanism leading to the violation of the orthogonal spin-momentum locking and the absence of any one of the three factors restores the spin-momentum orthogonality. In this work, we promote this idea into a dynamical framework of the interacting surface spin. Furthermore, we apply the emergent theory to the specific EPI with the $7 meV$ optical phonon mode\cite{Batanouny1,Batanouny2}. Our work therefore builts a theoretical bridge between the electron-optical phonon coupling experiments in Ref's\cite{Batanouny1,Batanouny2} and the in-plane spin anomaly experiments in Ref.\cite{Y.H.Wang}.       

In Section-II, we develop the Green's function theory of the interacting spin. In Section-III, we start with a microscopic electron-phonon interaction model and derive a set of equations for the spin dependent components of the SE. In section-III-B a convenient parametrization is introduced for these set of equations. Using experimental results in Ref's \cite{Batanouny1,Batanouny2,Batanouny3,Batanouny4} this theory is applied on the specific optical EPI and the SE solutions are shown. In Section III-C a connection is built between the theory and the spin texture experiments where we discuss the importance of the spin measurement as a probe for interactions. In Section IV, we compare our theoretical results with the experiments in Ref.\cite{Y.H.Wang}. We finally discuss the puzzling results of the out-of-plane spin polarization in the context of our interacting theory. 

\section{II- Green's Function formalism for the interacting spin}
The $Bi_{2-y}Sb_ySe_xTe_{3-x}$ family is of primary importance among the TIs\cite{BiSe_exp,BiTe} with a single Dirac spectrum on each surface. This family has the $D_{3d}$ point group symmetry in the rhombohedral class. The unit cell has the quintuple layered structure coordinated hexagonally within each layer. In order to formulate the interacting surface states, we start with the full Hamiltonian ${\cal H}={\cal H}_0+{\cal H}_I$ where ${\cal H}_0$ is the independent particle and ${\cal H}_I$ is the interaction parts respectively. The ${\cal H}_I$ is a (spin-independent) interaction of which nature will be discussed in the next section. The ${\cal H}_0$ is given by ($\hbar=1$ from now on)
\be
{\cal H}_0=\sum_{\bs k} \hat{\Psi}^\dagger_{\bs k} \Bigl[ \xi_{\bs k}+{\mathfrak g}_{\bs k}.{\bs \sigma}  \Bigr] \hat{\Psi}_{\bs k} \label{hamilt01}
\ee
Here ${\bs k}=(k_x,k_y)=k(\cos\phi,\sin\phi)$ is the two-dimensional wavevector of the TSSs. The electronic bands are two-dimensional as demonstrated by the ARPES measurements\cite{BiSe_ARPES}. In Eq.(\ref{hamilt01}), $\hat{\Psi}^\dagger_{\bs k}=(\hat{e}^\dagger_{\uparrow \bs k} \,,\, \hat{e}^\dagger_{\downarrow \bs k})$ is the electron spinor, $\xi_{\bs k}=k^2/2m-\mu$ with $m$ as the electron band mass in the parabolic sector, $\mu$ is the chemical potential and ${\mathfrak g}_{\bs k}={\bs g}_{xy\bs k}+{\bs g}_{z\bs k}$ is the full spin-orbit vector representing the Dirac cone with ${\bs g}_{xy\bs k}=\alpha_0 k \hat{\bs e}_\phi$ as the Rashba type isotropic in-plane and ${\bs g}_{z\bs k}=\lambda_0 k^3 \cos(3\phi) \hat{\bs e}_z$  as the cubic-Dresselhaus type out-of-plane SOC components. The $\alpha_0$ and $\lambda_0$ are referred to as the Dirac velocity and the hexagonal warping (HW) strengths respectively of which strengths are extracted from the surface band measurements. Apart from the helical Dirac state represented by ${\bs g}_{xy\bs k}$, the existence of the ${\bs g}_{z\bs k}$ is allowed by the point group symmetry which is well established experimentally\cite{BiSe_exp,BiTe} and theoretically\cite{Fu_1} as responsible for the six-fold periodic Fermi surface anisotropy and the three-fold periodic out-of-plane spin polarization\cite{out-of-plane-polarization}. 

The spin texture ${\bs S}({\bs k})$ of the interacting topological surface is represented in terms of the electron Green's function (GF) as 

\be 
{\bs S}({\bs k})=\frac{1}{2}\lim_{\tau \to 0^+}{\it Tr} \Bigl\{i \underline{\cal G}^{\nu}({\bs k},\tau) \, \bs{\sigma} \Bigr\}
\label{spin1.0}
\ee
Here $\nu=+ (-)$ describe the upper (lower) spin-orbit eigen branches of the ${\cal H}_0$ and the ${\cal G}^{\nu}({\bs k},\tau)$ is the full interacting GF of the surface quasiparticles in the $2\times 2$ matrix form as    
\be
\underline{{\cal G}}^{\nu}({\bs k},\tau)=- T_\tau \langle \psi_0^\nu\vert {\hat S}(\infty,-\infty) \hat{\Psi}_{\bs k}(\tau) \hat{\Psi}_{\bs k}^\dagger (0) 
\vert \psi_0^\nu\rangle~.
\label{gf1}
\ee
The $\underline{{\cal G}}^{\nu}$ is calculated in the spin-orbit eigenstate  $\vert \psi_0^\nu\rangle$ of ${\cal H}_0$ and ${\hat S}(\infty,-\infty)$ is the S-matrix. From now on we will only study the spin in the $\nu=+$ branch and drop this index. The Eq.(\ref{spin1.0}) is most conveniently expressed in terms of the retarded GF $\underline{\tilde{\cal G}}({\bs k},\nu_-)$ which is obtained from the Matsubara Green's function $\underline{\tilde{\cal G}}({\bs K})=\int_0^\beta d\tau \, exp(i\nu_m \tau)\,\underline{{\cal G}}({\bs k},\tau)$ with ${\bs K}=({\bs k},i\nu_m)$, where $\nu_m=(2m+1)\pi/\beta$ are the Fermionic Matsubara frequencies, by $i\nu_m \to \nu_-=\nu-i\delta$ as \cite{Mahan}
\be
{\bs S}({\bs k})=\frac{1}{2}\lim_{\delta \to 0^+}\int \limits_{-\infty}^\infty \frac{d\nu}{2\pi}{\it Tr} \Bigl\{\underline{\tilde{\cal G}}({\bs k},\nu_-).{\bs \sigma}\Bigr\}
\label{spin1.2}
\ee

The Eq.(\ref{spin1.2}) is combined with the Dyson equation
\be
\underline{\tilde{\cal G}}^{-1}({\bs K})=\underline{\tilde{\cal G}}_{0}^{-1}({\bs K})-\underline{\Sigma}({\bs K})
\label{Dyson_eq}
\ee  
where $\underline{\tilde{\cal G}}$ is the interacting electron GF, 
\be
\underline{\tilde{\cal G}}_{0}({\bs K})=\frac{1}{(i\nu_m-\xi_{\bs k}) -{\mathfrak g}_{\bs k}.{\bs \sigma}}
\label{G_0.1}
\ee
is the noninteracting electron GF and
\be
\underline{\Sigma}({\bs K})=\Sigma_0({\bs K}) +{\bs\Sigma}({\bs K}).{\bs \sigma}
\label{Self-energy}
\ee
is the full electron SE with $\Sigma_0({\bs K})$ as the spin-neutral (SNSE) and ${\bs \Sigma}({\bs K})=(\Sigma_x,\Sigma_y,\Sigma_z)$ as the spin-dependent SE (SDSE) components. As earlier works on the self-energy concentrated largely on the SNSE, we ignore this part here and concentrate on the less studied SDSE. The $\underline{\tilde{\cal G}}({\bs K})$ in Eq.(\ref{Dyson_eq}) is represented as           
\be
\underline{\tilde{\cal G}}({\bs K})=\frac{1}{(i\nu_m-\xi_{\bs k}) -{\bs G}({\bs K}).{\bs \sigma}}
\label{gf_def3}
\ee 
The Eq.(\ref{gf_def3}) differs from Eq.(\ref{G_0.1}) by the replacement of ${\mathfrak g}_{\bs k}$ by ${\bs G}({\bs K})={\mathfrak g}_{\bs k}+{\bs \Sigma}({\bs K})$ as the result of Eq's (\ref{Dyson_eq}), (\ref{G_0.1}) and (\ref{Self-energy}). Once ${\bs \Sigma}({\bs K})$ is found using a microscopic model, Eq.(\ref{spin1.2}) is calculated as shown in the Appendix as 
\be
{\bs S}({\bs k})=\frac{1}{2}\hat{\bs G}^*({\bs k}) 
\label{spin_1.3}
\ee
with ${\bs G}^*({\bs k})={\mathfrak g}_{\bs k}+{\bs \Sigma}({\bs k},i\nu^*_{\bs k})$ representing the renormalized Dirac cone by the interactions at the physical pole position $\nu^*_{\bs k}$ of the full GF and $\hat{\bs G}^*={\bs G}^*/\vert {\bs G}^*\vert$. From now on we use a short notation ${\bs \Sigma}^*({\bs k})$ for ${\bs \Sigma}({\bs k},i\nu^*_{\bs k})$. The $\nu^*_{\bs k}$ is given for the $+$ branch as the solution of $i\nu^*_{\bs k}=\xi_{\bs k}+\vert {\mathfrak g}_{\bs k}+{\bs \Sigma}^*({\bs k})\vert$. The Eq.(\ref{spin_1.3}) is exact in closed form and reproduces the known results in various limits. In the noninteracting limit, it recovers the standart result 
\be
{\bs S}_0({\bs k})=\frac{1}{2}\,{\hat {\mathfrak g}}_{\bs k}
\label{noninteracting_spin}
\ee
 where $\hat{\mathfrak g}_{\bs k}={\mathfrak g}_{\bs k}/\vert {\mathfrak g}_{\bs k} \vert$ and the perpendicular spin-momentum locking of the in-plane spin. Eq.(\ref{spin_1.3}) also correctly yields the Hartree-Fock limit\cite{TH_STA}.  

The three unit vectors $\hat{\bs g}_{xy\bs k}, \hat{\bs k},\hat{\bs z}$ form a convenient natural basis in which we represent ${\bs \Sigma}^*({\bs k})$ in terms of its components along and perpendicular to the $\hat{\bs g}_{xy\bs k}$ as 
\be
{\bs \Sigma}^*({\bs k})=\hat{\bs g}_{xy\bs k} ~ {\bs \Sigma}^*({\bs k}).\hat{\bs g}_{xy\bs k}+\hat{\bs g}_{xy\bs k}\times {\bs \Sigma}^*({\bs k})\times \hat{\bs g}_{xy\bs k}~ \nonumber \\.
\label{most_general_SE1}
\ee
The first term, ${\bs \Sigma}^*({\bs k}).\hat{\bs g}_{xy \bs k}$ is the regular in-plane component. The second term is a combination of the orthogonal components to $\hat{\bs g}_{xy\bs k}$ in which there is a regular out-of-plane component ${\Sigma}^*_{z\bs k}={\bs \Sigma}^*({\bs k}).\hat{\bs z}$ and finally ${\bs \Sigma}^*({\bs k}).\hat{\bs k}$ which we dub as the anomalous component. Here it will be shown that, this component arises in the presence of a Fermi surface anisotropy although the ${\mathfrak g}_{\bs k}$ has originally no such component, i.e. ${\mathfrak g}_{\bs k}.{\hat {\bs k}}=0$.  

The effect of the interactions on the spin is determined by the difference between the Eq's(\ref{spin_1.3}) and (\ref{noninteracting_spin}) as 
\be
\Delta {\bs S}({\bs k})=\frac{1}{2} 
[\hat{\bs G}^*({\bs k})-\hat{\mathfrak g}_{\bs k}] 
\label{spin_1.4}
\ee  
It is clear from Eq.(\ref{spin_1.4}) that the leading contribution to $\Delta {\bs S}({\bs k})$ is perpendicular to ${\bs S}_0({\bs k})$. When the interactions are much weaker than the SOC, $\vert {\bs \Sigma}({\bs k,\omega^*_{\bs k}}) \vert \ll \vert {\mathfrak g}_{\bs k}\vert $, the leading contribution to $\Delta {\bs S}({\bs k})$  
is given by the second term in Eq.(\ref{most_general_SE1})    
\be
\Delta {\bs S}({\bs k}) \simeq \frac{1}{2} \, \frac{\hat{\bs g}_{xy\bs k}\times {\bs \Sigma}^*({\bs k})\times \hat{\bs g}_{xy\bs k}}{\vert {\mathfrak g}_{\bs k}\vert}  
\label{spin_1.5}
\ee 

This result shows that, the interactions can cause directional spin anomalies. However, interaction per se, is not sufficient to produce a change perpendicular to the original direction of the spin. Indeed, if the full Hamiltonian ${\cal H}$ is invariant under continuous in-plane rotations, the ${\bs g}_{z\bs k}=0$ and the ${\bs \Sigma}^*({\bs k})$ and the resulting ${\bs G}^*({\bs k})$ remain parallel to ${\bs g}_{xy\bs k}$  
and no anomalous component is produced. This already hints that a lower symmetry than continuous rotations is essential to produce a finite $\Delta {\bs S}({\bs k})$ in Eq.(\ref{spin_1.5}). 

We now represent the in-plane components of ${\bs \Sigma}^*({\bs k})$ in the complex form as 
$\Sigma^*_{xy\bs k}=\Sigma^*_{x\bs k}-i\Sigma^*_{y\bs k}=e^{i\phi_S} T_{\bs k}$, where $T_{\bs k}=T_{R\bs k}+iT_{I\bs k}$ is an even function by the TRI and is allowed to be complex. Here the $T_{R\bs k}$ is the regular component, whereas the $T_{I\bs k}$ is connected with the anomalous component of the spin. They can be combined in the suggestive form ${\bs \Sigma}^*_{xy\bs k}= T_{R\bs k}\, \hat{\bs g}_{xy\bs k}+T_{I\bs k}\, \hat{\bs k}$. This complex form will be  particularly useful when the real crystal symmetries are considered in Section.III. In this notation, the leading term in Eq.(\ref{spin_1.5}) is  
\be
\Delta {\bs S}({\bs k})\simeq \frac{1}{2} \, \frac{T_{I\bs k} \, \hat{\bs k} +\Sigma^*_{z \bs k}}{\vert {\mathfrak g}_{\bs k}\vert}  
\label{spin_1.6}
\ee    
The Eq.(\ref{spin_1.6}) builts the connection between the spin texture anomalies and the interaction. We now device a model for microscopic interactions in order to calculate the right hand side of Eq.(\ref{spin_1.6}).  

\section{III- A Model for interactions in the spin-dependent channel} 

In the previous section, we established the theoretical connection between the experimental STA and the imaginary part $T_{I\bs k}$ of the off-diagonal SDSE. We now consider the recent experiments\cite{Batanouny1,Batanouny2,Batanouny3,Batanouny4} in $Bi_2(Se_xTe_{3-x})$ where a strong e-ph coupling was found between the topological surface electrons and an optical phonon mode near the $7 meV$ energy region. Our aim in this article is to find the contributions of this mode to the SDSE in a material based approach using the interaction parameters found therein. Finally we connect it to the $\Delta {\bs S}({\bs k})$ in Eq.(\ref{spin_1.6}) and  discuss the results together with two other experiments\cite{Y.H.Wang,Nomura} for the in-plane and out-of plane STAs. 

The ${\cal H}_I$ is now considered to be the EPI between the topological surface electrons and the phonons given by 
\be
{\cal H}_{I}=\sum_{{\bs k},{\bs k^\prime}}\rho_\gamma({\bs k},{\bs k^\prime}){\hat A}_\gamma({\bs k}-{\bs k^\prime})\hat{\Psi}^\dagger_{\bs k}\sigma_z\hat{\Psi}_{\bs k^\prime}+h.c.~~~~~
\label{hamilt_int}
\ee 
The ${\hat A}_\gamma({\bs Q})=({\hat a}_{\gamma\bs Q}+{\hat a}^\dagger_{\gamma \,-\bs Q})$ is the phonon displacement operator in terms of the phonon mode annihilation/creation operators  ${\hat a}_{\gamma\bs Q}, {\hat a}^\dagger_{\gamma\bs Q}$. The phonon branch denoted by $\gamma$ will be considered as the $7 meV$ optical mode observed in Ref.'s \cite{Batanouny1,Batanouny2,Batanouny3,Batanouny4}. Since we will be considering only one phonon mode, the $\gamma$ index will be dropped from here on. The EPI is given by 
\be
\rho({\bs k},{\bs k^\prime})=\lambda({\bs k}-{\bs k^\prime}) \,{\bs k^\prime}\times {\bs k}.\hat{e}_z
\label{general_e-ph_coupling}
\ee 
where $\lambda({\bs k}-{\bs k^\prime})$ is the microscopic coupling constant, and the ${\bs k^\prime}\times {\bs k}.\hat{e}_z \propto \sin(\phi-\phi^\prime)$ is the structure factor as derived by Thalmeier\cite{Thalmeier} that arises when the initial and the final electron states are on the surface. The additional $\sin(\phi-\phi^\prime)$ factor inhibits the complete backward scattering as required by the time-reversal symmetry, while permiting scattering at finite angles.  

\subsection{A-The calculation of the SDSE} 
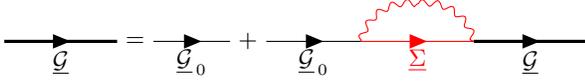
\begin{figure} 
\begin{tikzpicture}
\begin{feynman}
\vertex(v0);
\vertex[right=1.5cm of v0](v1){\(=\)}; 
\vertex[right=1.5cm of v1](a1){\(+\)}; 
\vertex[right=1.5cm of a1] (a2);
\vertex[right=1.5cm of a2] (a3);
\vertex[right=1.5cm of a3] (a4);
\vertex at ($(a2)!0.5!(a3)!0.5cm!90:(a3)$) (d);
\diagram* {
(v0) -- [very thick,fermion,edge label'=\(\underline{\cal G}\)] (v1) -- [fermion,edge label'=\(\underline{\cal G}_{\,0}\)] (a1)-- [fermion,edge label'=\(\underline{\cal G}_{\,0}\)] (a2) -- [red,fermion,edge label'=\(\underline{\Sigma}\)] (a3) -- [very thick,fermion,edge label'=\(\underline{\cal G}\)] (a4),
(a2)-- [red,boson,quarter left] (d) -- [red,boson,quarter left] (a3),
};
\end{feynman}
\end{tikzpicture}
\caption{The Feynman diagram for the Dyson equation represented by Eq.(\ref{Dyson_eq}).}
\label{dyson_diagram}
\end{figure}

The ${\bs \Sigma}({\bs K})$ is calculated starting from the fundamental interaction vertex in Eq.(\ref{hamilt_int}) as shown in the Fig.(\ref{dyson_diagram}). 
%%%%%%%%%%%%%%%%%%%%%%%%%%%%%%%%%%%%%%%%%%%%%%%%%%%%%%%%%%%%%%%%%%%%%%%%%%%%%%%%%%%%%%%%%%%%%%%%%%%%%%%%
%%%%%%%%%%%%%%%%%%%%%%%%%%%%%%%%%%%%%%%%%%%%%%%%%%%%%%%%%%%%%%%%%%%%%%%%%%%%%%%%%%%%%%%%%%%%%%%%%%%%%%%%%%%%%%%
Using the Hartree-Fock-Migdal approximation the left hand side of Eq.(\ref{Self-energy}) is\cite{Grimvall,Mahan}
\be 
\underline{\Sigma}({\bs K})&=&-\frac{1}{\beta}\sum_{{\bs K}^\prime}
\vert \rho({\bs k},{\bs k^\prime})\vert^2 \,[\sigma_z \,{\underline{\tilde{\cal G}}}({\bs K^\prime}) \,\sigma_z] \,D_0({\bs K}-{\bs K^\prime}) \nonumber\\
\label{self-energy-matrix}
\ee
where ${\bs K}^\prime=({\bs k}^\prime,i\nu_{m^\prime})$ and $D_0({\bs q},i\omega_{mm^\prime})=-2\Omega_{\bs q}/(\omega^2_{mm^\prime}+\Omega^2_{\bs q})$ is the free phonon propagator with ${\bs q}={\bs k}-{\bs k^\prime}$ and $\omega_{mm^\prime}=\nu_m-\nu_{m^\prime}$.  
We now concentrate on the self-energy corrections on the $+$ bare spin-orbit branch $E_{\bs k}=\xi_{\bs k}+ \vert {\mathfrak g}_{\bs k}\vert$. Performing the Matsubara frequency summation in Eq.(\ref{self-energy-matrix}) on $\nu_{m^\prime}$,
\be
{\bs \Sigma}({\bs K})&=&-\frac{1}{2}\sum_{\bs k^\prime} \vert \rho(\bs k, \bs k^\prime)\vert^2 \hat{{\bs \gamma}}({\bs K^\prime}) \, \chi({\bs k},{\bs k^\prime}) 
\label{sigma_matrix_1}
\ee
where $\hat{\bs \gamma}=(-G_x,-G_y,G_z)/\vert {\bs G}\vert$ is the same as $\hat{\bs G}$ but rotated along the z-direction by $\pi$. This form arises due to the specific dependence of the EPI in Eq.(\ref{hamilt_int}) on $\sigma_z$. The last quantity in (\ref{sigma_matrix_1}) is       
\be
\chi&=&\frac{n_B(\Omega_{\bs q})+n_F(E_{\bs k^\prime})}{i\nu_m-E_{\bs k^\prime}+\Omega_{\bs q}}+\frac{n_B(\Omega_{\bs q})+1-n_F(E_{\bs k^\prime})}{i\nu_m-E_{ \bs k^\prime}-\Omega_{\bs q}}  ~.
\label{chi}
\ee 
Eq.(\ref{sigma_matrix_1}) has the natural outcome that, if the system has continuous rotational symmetry in the surface plane, the ${\bs \Sigma}({\bs K})$ has only a non-vanishing in-plane component parallel to $\hat{\bs g}_{xy\bs k}$. 

In order to connect Eq.(\ref{sigma_matrix_1}) to the spin, the ${\bs \Sigma}({\bs K})$ must be converted to the retarded SDSE ${\bs \Sigma}({\bs k},\nu-i\delta)$ and then calculated at the physical pole position $\nu =\nu^*_{\bs k}$ as outlined below Eq.(\ref{spin_1.3}) and shown in the Appendix. The Eq.(\ref{sigma_matrix_1}) is then fully determined by the electronic degrees of freedom, the phonon energy and the EPI constant as 
\be
{\bs \Sigma}^*({\bs k})=-\frac{1}{2}\sum_{\bs k^\prime} V_{\mathscr{F}}({\bs k},{\bs k^\prime}) {\hat {\bs G}^*({\bs k^\prime})} 
\label{tetragonal_sigma_3}
\ee 
where $V_{\mathscr{F}}({\bs k},{\bs k^\prime})=V({\bs k},{\bs k^\prime}) \mathscr{F}_{\bs k^\prime}$  with $\mathscr{F}_{\bs k^\prime}=\tanh(\beta \xi_{\bs k^\prime}/2)$ and an effective interaction potential 
\be
V({\bs k},{\bs k^\prime})=\frac{\vert \rho({\bs k},{\bs k^\prime})\vert^2}{\Omega_{{\bs k}-{\bs k^\prime}}}~.
\label{eff_pot}
\ee  
The Eq.(\ref{tetragonal_sigma_3}) is finally written in the ${\hat{\bs g}}_{xy\bs k},\hat{\bs k},\hat{\bs z}$ basis as a self-consistent set of equations  
\be
T_{R\bs k}&=&\sum_{\bs k^\prime} \frac{V_{\mathscr{F}}}{\vert {\bs G}_{\bs k^\prime} \vert}\, \Bigl({\widehat{\bs g \bs g^\prime} \, [g_{xy\bs k^\prime}+T_{R\bs k^\prime}]}+{\widehat{\bs g \bs k^\prime}} \, T_{I\bs k^\prime}\Bigr) ~~~~ \label{polar_notation_1} \\
T_{I\bs k}&=&\sum_{\bs k^\prime} \frac{V_{\mathscr{F}}}{\vert {\bs G}_{\bs k^\prime} \vert} \Bigl( {\widehat{\bs k \bs g^\prime} \, [g_{xy\bs k^\prime}+T_{R\bs k^\prime}]}+{\widehat{\bs k \bs k^\prime}} \, T_{I\bs k^\prime} \Bigr) ~~~~ \label{polar_notation_2} \\
\Sigma^*_{z\bs k}&=&-\sum_{\bs k^\prime} \frac{V_{\mathscr{F}}}{\vert {\bs G}_{\bs k^\prime} \vert} 
[{g_{z \bs k^\prime}+\Sigma^*_{z\bs k^\prime}]} ~~~~ \label{polar_notation_3}
\ee
where we defined ${\widehat{\bs g \bs g^\prime}}={\widehat{\bs k \bs k^\prime}}=cos(\phi-\phi^\prime)$ and ${\widehat{\bs g \bs k^\prime}}=-{\widehat{\bs k \bs g^\prime}}=\sin(\phi-\phi^\prime)$.

The input to the Eq's(\ref{polar_notation_1}-\ref{polar_notation_3}) is the IPP parameters of ${\cal H}_0$ and those of the interaction Hamiltonian given in Eq.(\ref{eff_pot}). The former can be deduced from the spectral measurements of the surface state. It is difficult to have a resonable estimate of the latter parameters directly from a microscopic theory without a phenomenological support from experiments. Both parameters were also extensively studied using different approaches from the phonon\cite{Batanouny2} and the electron\cite{Batanouny3} degrees of freedom. A detailed ab-initio study was also recently presented in Ref.\cite{Heid}. 

We introduce below a convenient parametrization to solve Eq's(\ref{polar_notation_1}-\ref{polar_notation_3}). With the phenomenological input provided by the experiments, we then present the solution.  

\subsection{B-Parametrization and solution of the SDSE}
\begin{figure}[h] 
% gplt: sanga/sanga_gplt  
% data: veri dosyası değil
\includegraphics[scale=0.32]{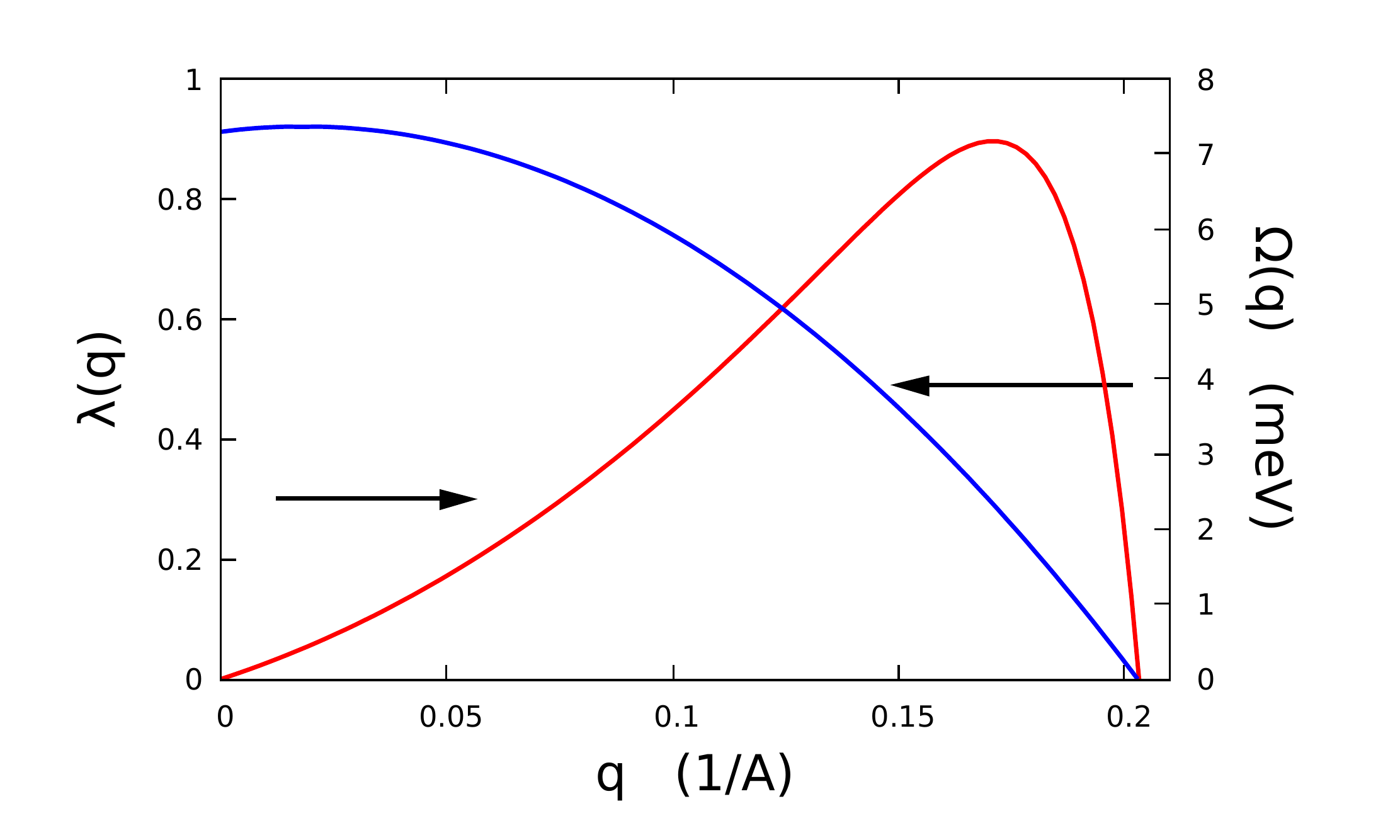}    
\vskip-0.3truecm
\caption{The phonon frequency and the interaction strength in Ref.\cite{Batanouny3} for $Bi_2Se_3$. (The data is provided by M. El-Batanouny)}
\label{e-ph_parameters}
\vskip-0.3truecm
\end{figure}

The components of ${\bs \Sigma}({\bs k})$ transform under the representations of the crystallographic point group. For the $Bi_{2-y}Sb_ySe_{3-x}Te_x$ family, the complex and even $T_{\bs k}$ is invariant under in-plane rotations by $2\pi/6$ whereas the real and odd $\Sigma^*_{z\bs k}$ is invariant under the $2\pi/3$ in-plane rotations. We expand in their relevant symmetry basis,   
\be
T_{\bs k}&=&\sum_{m=-\infty}^{\infty}\,\mathscr{T}_{m k}\,e^{i 6 m \phi} 
\label{2D_sigma_z_T_k_1_b}
\\
\Sigma^*_{z\bs k}&=& 
\sum_{m=0}^{\infty}\,\mathscr{S}_{m k} \cos[3(2m+1)\phi] 
\label{2D_sigma_z_T_k_1_a} 
\ee
The solution of the Eq's(\ref{polar_notation_1}-\ref{polar_notation_3}) is then converted to finding the complex $\mathscr{T}_{m k}$ and the real $\mathscr{S}_{m k}$ coefficients which are radial functions of $k$. For numerical solution, this infinite set is truncated in this work beyond the leading terms $\mathscr{T}_{0 k}, \mathscr{T}_{\pm 1 k}$ and $\mathscr{S}_{0 k}$ respectively with the assumption that the neglected terms corresponding to the higher symmetry harmonics are negligible. We adopt the electron, phonon and the interaction parameters using the available experimental results in Ref.\cite{Batanouny3} for $Bi_2Se_3$ as shown in Fig.(\ref{e-ph_parameters}). The mode-specific phonon properties for other compounds in the family would be ideal to study the spin anomaly in the whole compound but we are currently unaware of such data.    
The numerical solution of the dimensionless $T_{\bs k}$ and $\Sigma^*_{z\bs k}$ for  $Bi_2Se_3$ scaled by the $E_F\simeq 0.3 eV$ are shown in Fig.(\ref{SE_xyz_solution}) along the $\bar{\Gamma}\bar{K}$ direction. The $k$ axis is made dimensionless by the natural length scale $d_0=\sqrt{\lambda_0/\alpha_0} \simeq 15 \AA$ for $Bi_2Se_3$. On the second horizontal axes we use the dimensionless HW strength $\bar{\lambda}_0=\lambda_0/(E_F d^3)$ where $E_F \simeq 0.3 eV$ is the Fermi energy in $Bi_2Se_3$ with respect to the $\Gamma$ point.      

We note that the regular solution $T_{R\bs k}$ survives in the absence of the warping anisotropy. By the Eq.(\ref{2D_sigma_z_T_k_1_b}), 
\be
T_{R\bs k}\simeq \mathscr{T}_{0k}+(\mathscr{T}_{1k}+\mathscr{T}_{-1k}) \cos6\phi ~.
\label{regular_sol}
\ee
Here, $\mathscr{T}_{0k}$ is an isotropic renormalization for the linear spectrum (for further discussion see Section IV.D). The correction given by the second term in Eq.(\ref{regular_sol}) does not create a significant anisotropy in the spectrum. The anomalous component $T_{I\bs k}$ and the $\Sigma^*_{z\bs k}$ increase with increasing hexagonal anisotropy. The overall picture is than consistent with that, the HW is an important factor in the spin direction. By the Eq.(\ref{spin_1.6}), the ratio $T_{I\bs k}/\vert {\mathfrak g}_{\bs k}\vert$ is a measure of the in-plane deviation from the orthogonal spin-momentum locking. A quick estimate of the deviation can be made using the middle plot in Fig.(\ref{SE_xyz_solution}). At the $k_F \simeq 0.1 \AA^{-1}$, which corresponds to the central part of the $kd$ axis, $T_{I\bs k}\simeq 0.01 E_F\simeq 3 meV$. Using the Fermi velocity for $Bi_2Se_3$, the amount of in-plane deviation is found to be close to $10$ degrees. We will examine $T_{I\bs k}$ in more detail in the Section V below. Finally, the $\Sigma_{z\bs k}$ is shown in the right plot in Fig.(\ref{SE_xyz_solution}) showing that a significant out-of-plane renormalization can be produced by the HW.  

As a general feature of the solution summarized in three plots for $T_{R\bs k},T_{I\bs k}, \Sigma_{z\bs k}$ in Fig.(\ref{SE_xyz_solution}), the interaction effects due to the SDSE are marginal in the large momentum and large HW region whereas they are almost negligible at and below the $E_F$. The left plot in Fig.(\ref{SE_xyz_solution}) should be considered with the aid of Fig.(\ref{renormalization_Dirac_spectrum}) where the contribution of the $7meV$ phonon mode in the shape of the Dirac cone is summarized. There, the left plot is for the extracted IPP parameters in $Bi_2Se_3$ whereas the right plot is when we turned off the HW. Relatively stronger contributions are found near the Fermi surface where a strong asymmetry is introduced by the presence of warping. 

The Fig.(\ref{SE_xyz_solution}) demonstrates that the coupling to this specific phonon mode has almost no effect in a large part of the surface electronic spectrum. This observation agrees with the earlier experiments and shows that the SNSE is the major source for the band renormalization. This leaves the extraction of the SDSE a challenge for experimentalists, but the situation is not completely hopeless. With the progress in the high precision spin and energy resolved measurements, it is possible to extract the full spin-texture ${\bs S}({\bs k})$ from which useful information about the SDSE can be collected as we discuss next. 

\subsection{C- Spin as a probe for interactions}
The SDSE can be indirectly probed by measuring the ${\bs S}({\bs k})$. The spin-ARPES and the Scanning Tunnelling  
\newpage
\onecolumngrid

\begin{figure}[h] 
% gplt: fig_SE_xyz_onecolumn_gplt  
% data: SE_xyz_short_Batanouny_socz_pio12
\includegraphics[scale=0.27]{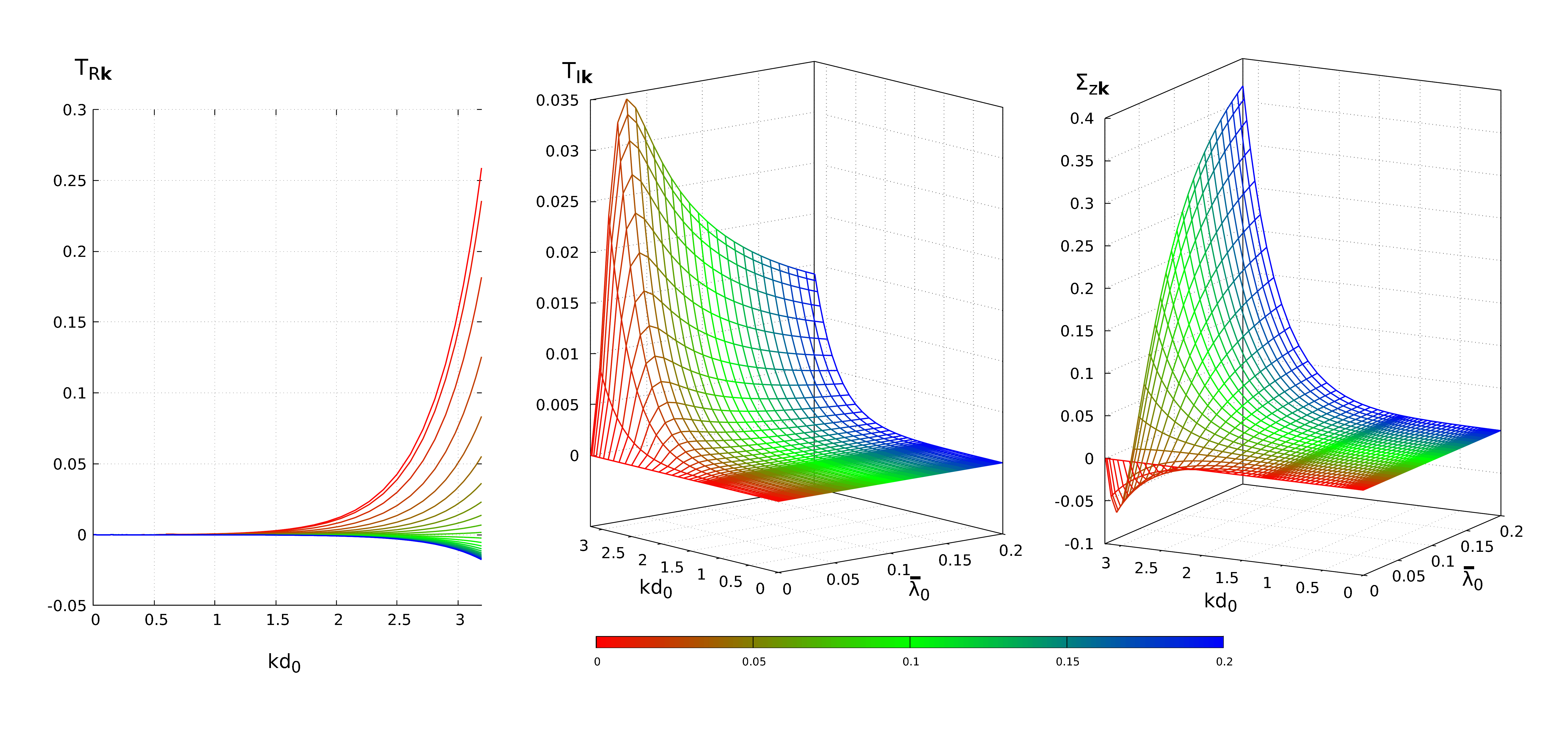}    
\vskip-0.3truecm
\caption{The components of the ${\bs \Sigma}^*_{\bs k}$ in the $\{\hat{\bs g}_{xy\bs k}, \hat{\bs k},\hat{\bs z}\}$ basis as written in Eq's (\ref{polar_notation_1}-\ref{polar_notation_3}) along the $\bar{\Gamma}\bar{M}$ line for the dimensionless HW coupling range $0 \le \bar{\lambda_0} \le 0.2$ [where $\bar{\lambda_0}=\lambda_0/(E_F d^3)$] and the dimensionless momentum $kd_0$ [where $d_0=\sqrt{\lambda_0/\alpha_0}$] (The colorbox represents the HW coupling strength as an additional guide to the eye). All numerical values on the vertical axes are scaled by $E_F$.} 
\label{SE_xyz_solution}
%\vskip-0.3truecm
\end{figure}
\twocolumngrid

Spectroscopy are most widely used techniques in TI experiments where the in-plane spin and momentum locking was first demonstrated. The high resolution time-of-flight-spin-ARPES with a circularly polarized light source has also been applied successfully\cite{Y.H.Wang}. Using Eq.(\ref{spin_1.6}) and ignoring the second order terms in the interaction, the in-plane spin is  
${\bs S}({\bs k}) .\hat{\bs k} \simeq (1/2) \, T_{I\bs k}/\vert {\mathfrak g}_{\bs k}\vert$ 
the left hand side of which is directly measurable. The right hand side is obtained from the theory in Section.II. Indeed, $T_{I\bs k}$ has an isotropic contribution from the imaginary part of $\mathscr{T}_{0 k}$ and an anisotropic contribution from $\mathscr{T}_{\pm 1 k}$ which goes like $\sin(6\phi)$. We therefore derive an expression for $\mathscr{T}_{m k}$ by inverting Eq.(\ref{2D_sigma_z_T_k_1_b}) and using (\ref{polar_notation_2}) as 
\be
\mathscr{T}_{m k}=
\sum_{k^\prime} v_m(k,k^\prime) \mathbb{G}_{m k^\prime}
\label{mathscrtmks}
\ee
where $\mathbb{G}_{m k^\prime}=\langle e^{-i 6m \phi^\prime}{\mathscr G}_{xy}({\bs k^\prime})\rangle_{\phi^\prime}$ with $\langle \dots \rangle_{\phi^\prime}$ indicating the angular average and  
\be
{\mathscr G}_{xy}({\bs k^\prime})=\frac{g_{xy\bs k^\prime}+T_{\bs k^\prime}}{\vert \bs G({\bs k^\prime})\vert} \mathscr{F}_{\bs k^\prime}
\label{mathscrtmks2}
\ee
The $v_m(k,k^\prime)$ in Eq.(\ref{mathscrtmks}) is the $6m-1$'th angular momentum component of the interaction potential   
\be
v_m(k,k^\prime)=\langle V({\bs k},{\bs k^\prime}) \,e^{-i(6m-1)(\phi-\phi^\prime)} \rangle_{\phi-\phi^\prime} 
\label{v_m}
\ee
Knowing that $T_{I\bs k}$ is generated by the $\mathscr{T}_{m k}$ and $\mathscr{T}^*_{-m k}$, we concentrate on the $v_m$ and $v^*_{-m}$ in Eq.(\ref{mathscrtmks}). These are induced by different angular momentum components as given by the Eq.(\ref{v_m}) and generally unequal. For $m=\pm 1$ these angular momentum components have quantum numbers $-5$ and $7$. Hence the $\mathscr{T}_{1 k}$ and $\mathscr{T}^*_{-1 k}$ are unequal for the general case and this drives the leading term in the six-fold symmetric contribution in $T_{I\bs k}$. The in-plane spin is given by 
\be
{\bs S}_{xy\bs k}=\frac{1}{2} \frac{(\alpha_0 k+ T_{R\bs k})\,\hat{\bs g}_{\bs k}+T_{I\bs k}\,\hat{\bs k}}{\vert {\bs G}_{\bs k}\vert}~. 
\label{full_spin_1}
\ee 
We now introduce the generalized symmetry moments $\tau_{mk}=(\mathscr{T}_{m k}-\mathscr{T}^*_{-m k})/2i$. Using Eq.(\ref{2D_sigma_z_T_k_1_b}) in  (\ref{spin_1.6}) we have   
\be
\frac{\tau_{mk}}{\vert {\bs G}_{\bs k} \vert} = \frac{2}{1}\langle {\bs S}_{xy\bs k} \, .\,\hat{\bs k} \,e^{i 6m  \phi}\rangle_\phi 
\label{spin_text_2}
\ee
The $\tau_{mk}$'s are generally complex. The in-plane spin-texture can be reconstructed from the inverse of Eq.(\ref{spin_text_2}) as 
\be
{\bs S}_{xy\bs k}.\hat{\bs k}=\frac{1}{2}\sum_{m=0}^\infty \frac{\vert \tau_{mk}\vert}{\vert {\bs G}_{\bs k}\vert}\,\sin(6m\phi+\Lambda_m)
\label{n_odd_3.1}
\ee
where
\be
\vert \tau_{mk}\vert=\sqrt{\operatorname{Re}\{\tau_{mk}\}^2+\operatorname{Im}\{\tau_{mk}\}^2} 
\label{tau_mk_mag}
\ee
and 
\be
\Lambda_m=\tan^{-1} \Bigl(\operatorname{Re}\{\tau_{mk}\} / \operatorname{Im}\{\tau_{mk}\}\Bigr)
\label{tau_mk_phase}
\ee
where the latter is a reference angle for the 6m-folded component of the anomalous spin. The $\Lambda_m$ is revealed easily by the measurement of the spin at high symmetry directions. For $Bi_2Se_3$ we consider the $m=\pm 1$ term in Eq.(\ref{n_odd_3.1}) and the spin deviation vanishes at the high symmetry directions $\bar{\Gamma}\bar{M}$ and $\bar{\Gamma}\bar{K}$\cite{Y.H.Wang}. This yields $\Lambda_1=0$ or $\pi$. In $Bi_2Se_3$ the spin is canted outward between $\bar{\Gamma}\bar{M}$ and $\bar{\Gamma}\bar{K}$ which further implies that $\Lambda_1=0$.

\begin{figure}[h] 
% produced by gplt_dirac_cone_h_dene in  /home_research/trsb/bisete2/yaz_2018/BCK_TH_colobration/hexagonal_warping/July_19/seud/seud_complex_parallel
% data file dene1_high.2_socz   and  dene1_pio2_high.2_socz    
\includegraphics[scale=0.37]{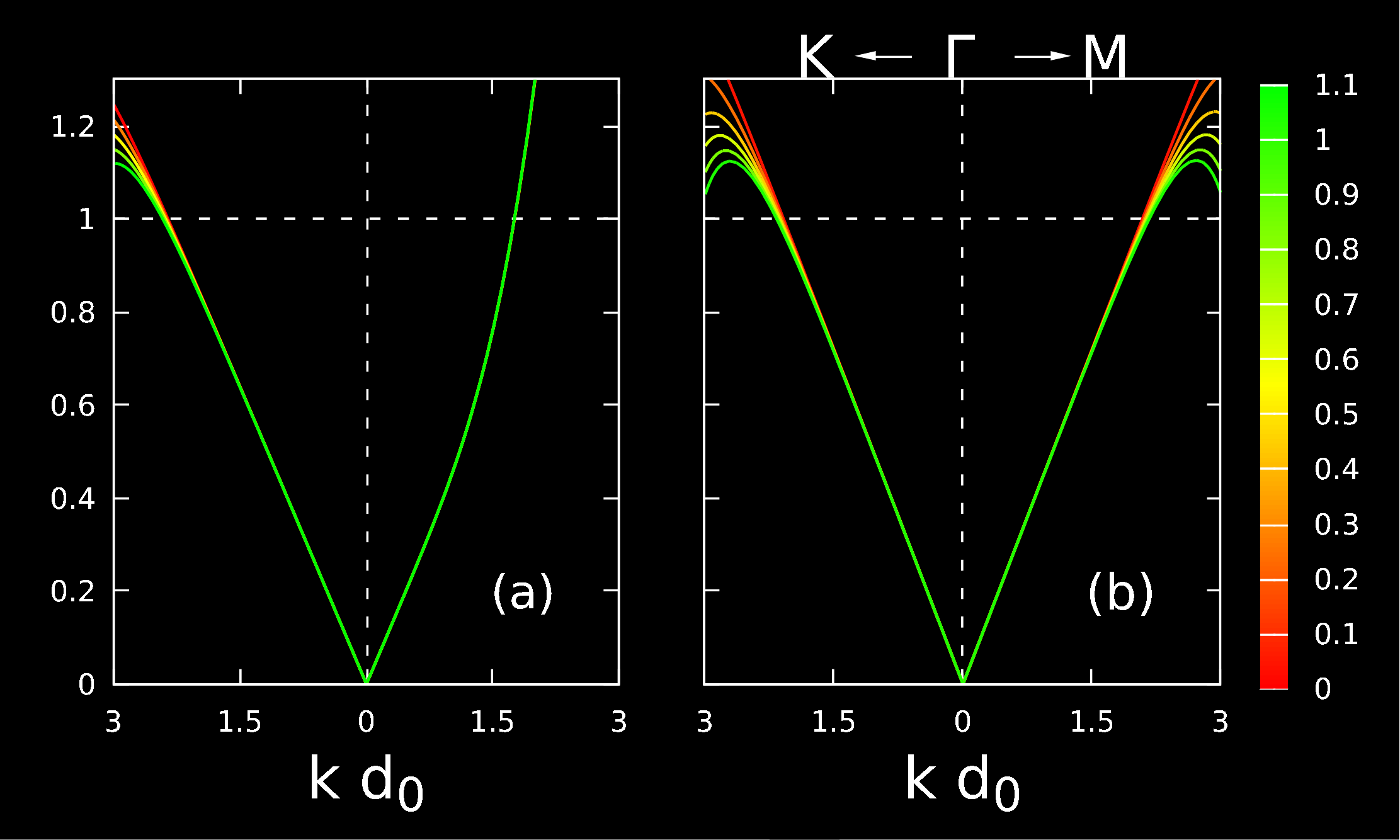}   
\vskip-0.3truecm
\centering
\caption{The renormalization of the Dirac spectrum as a result of the Eq's(\ref{polar_notation_1}-\ref{polar_notation_3}) along the $\bar{\Gamma} \bar{M}$ and $\bar{\Gamma} \bar{K}$ directions and for different interaction strengths $0 \le f_0 \le 1.1$ as shown on the right color scale. The vertical axis is energy normalized by $E_F$. a) Here, all free particle parameters are adopted for $Bi_2Se_3$. The $f_0=1$ corresponds to the interaction strength measured for this compound in Ref.\cite{Batanouny3}, b) The same as in the left plot without hexagonal warping $\lambda_0=0$.}
\label{renormalization_Dirac_spectrum}
\vskip-0.3truecm
\end{figure}

A few words about the isotropic contribution to the $T_{\bs k}$ can be made here. 
If the continuous rotational symmetry is manifest in the full Hamiltonian, $\Sigma_{z\bs k}$ vanishes. All $\mathscr{T}_{m k}$'s with $m\ne 0$ also vanish. In the theory here, the $\mathscr{T}_{0 k}$ is allowed to be complex with a spontaneously arising phase under specific interactions. The experiment in Ref.\cite{Y.H.Wang} indicates that $\mathscr{T}_{0 k}$ is real in the $Bi_2Se_3$. A real $\mathscr{T}_{0 k}$ is can be relevant in the renormalization of the bare linear Dirac spectrum. Expanding as 
\be
\mathscr{T}_{0 k}=k(t_0+t_1 k+ t_2 k^2 +\dots)
\label{exp_t0k}
\ee
the effective spin-orbit contribution becomes $g_{xy k}+\mathscr{T}_{0 k}=k v_k$ with a renormalized slope $\alpha_0$ at the $\Gamma$ point and the nonlinear velocity profile $v_k=\alpha_0+t_0+t_1 k+ t_2 k^2 +\dots$. Similar corrections have been suggested for $Bi_2Se_3$\cite{Y.H.Wang} and for $Bi_2Te_3$\cite{Basak} phenomenologically. The Fig.(\ref{renormalization_Dirac_spectrum}) summarizes the renormalization effects of the Dirac spectrum by  interactions. In the calculations leading to this plot, we introduced a multiplicative interaction strength $f_0$ by the substitution $V({\bs k},{\bs k^\prime}) \to f_0 \, V({\bs k},{\bs k^\prime})$ in Eq.(\ref{eff_pot}). The colors in the plots refer to different values of $f_0$ in the range $0 \le f_0 \le 1.1$ introduced to qualitatively understand the role played by different coupling strengths. For $f_0=1$ the coupling strength corresponds to the EPI in $Bi_2Se_3$ studied in Ref.\cite{Batanouny3} and shown in Fig.(\ref{e-ph_parameters}). The left plot in Fig.(\ref{SE_xyz_solution}) indicates that there is almost no renormalization of the Dirac velocity in a large range around the $\Gamma$ point. The change is stronger near $E_F$ but the overall deviation from the original spectrum is negligible. 
   
In this section, we have the conclusion that the effect of the SDSE in the energy spectrum is negligible whereas, it has direct connection with the anomalies in the in-plane spin direction away from the $\Gamma$-point. This result is consistent with the great success of the IPP in the vicinity of the $\Gamma$ point. We now examine the $7 meV$ phonon mode and its coupling to the surface states for driving the spin anomalies. As a side remark, we also comment in this context, on the out-of-plane spin polarization. 

\section{IV- Can the e-ph interaction be the origin for spin texture anomaly?}

It is shown here that the microscopic interaction of the $7meV$ optical surface phonon mode with the surface electrons, is a mechanism leading to the experimentally observed STA. Before this is shown, we demonstrate how the mode-resolved dimensionless EPI constant $\lambda_{\bs q}$ in Ref.\cite{Batanouny2} is used to find the effective interaction in Eq.(\ref{eff_pot}). 
In the electron propagator formalism of the Eliashberg theory, the mode-resolved e-ph coupling constant is defined by\cite{Grimvall} 
\be
\lambda_{\bs q}= {\cal N}(E_F) \,\langle \langle V({\bs k},{\bs k^\prime})\rangle \rangle 
\label{mode_resolved_lambda}
\ee 
where ${\cal N}(E_F)$ is the electron density of states at the Fermi level and $\langle \langle \dots \rangle \rangle$ is the Fermi surface average. In the phonon propagator based approach in Ref.\cite{Batanouny2} it was considered to be  
\be
\lambda_{\bs q}=\frac{\gamma_{\bs q}}{\pi{\cal N}(E_F) \Omega^2_{\bs q}}
\label{e-ph_Batanouny_eq1}
\ee     
where ${\cal N}(E_F)=D_0 k_F/(k_F+2\pi D_0 \alpha_0)$ is the  density of surface states at the Fermi energy with $D_0=m/(2\pi)$ and $\alpha_0$ as define before, as the DOS of the two dimensional free electron gas. The $\gamma_{\bs q}$ is the imaginary part of the phonon self energy. 
At the lowest order the $\gamma_{\bs q}$ is given by\cite{PB-Allen}
\be
\gamma_{\bs q}&=&\pi \,\sum_{\bs k} \,\vert \rho({\bs k},{\bs k+ \bs q}) \vert^2\, \nonumber \\  
&\times& [n_F(\xi_{\bs k})-n_F(\xi_{{\bs k}+{\bs q}})]\,\delta(\Omega_{\bs q}-\xi_{{\bs k}+{\bs q}}+\xi_{\bs k}) \nonumber \\
\label{PBAllen_eq1}
\ee
here, different from the original reference in \cite{PB-Allen}, we replaced the microscopic e-ph coupling constant with the form in Eq.(\ref{general_e-ph_coupling}). The simplest microscopic estimate about $\lambda_{\bs q}$ is obtained at zero temperature for which the right hand side of the Eq.(\ref{PBAllen_eq1}) has contribution in a small energy window of size $\Omega_{\bs q}$ around the Fermi energy $E_F$. Also converting the sum into an integral over the density of states, we find in the vicinity $E\simeq E_F \ll \omega_{\bs q}$, 
\be
\lambda_{\bs q}={\cal N}(E_F)\,\frac{\vert \rho({\bs k},{\bs k}+{\bs q})\vert^2}{\Omega_{\bs q}}
\label{PBAllen_eq2}
\ee
Comparing Eq's(\ref{mode_resolved_lambda}), (\ref{PBAllen_eq2}) and (\ref{eff_pot}) we conclude that the electron based and the phonon based approaches consistently yield the same electron-phonon coupling constant. This implies that, we can directly use the e-ph coupling parameters in Ref.\cite{Batanouny1,Batanouny2,Batanouny3,Batanouny4} in the electron GF based approach combined in the form of an effective potential in Eq.(\ref{eff_pot}). The generalized form of the $\rho({\bs k},{\bs k}^\prime)$ in Eq.(\ref{general_e-ph_coupling}) accounts for the interaction of the phonons with the electrons on the topological surface, leaving only the magnitude of the coupling dependent on the material. We use this as a freedom to introduce a material dependent parameter $f_0$ which can account for different e-ph coupling strengths\cite{Chayou-Chen} in the family $Bi_{2-y} Sb_y Se_x Te_{3-x}$ as $V({\bs k},{\bs k^\prime}) \to f_0 V({\bs k},{\bs k^\prime})$ with the case $f_0=1$ corresponding to the $Bi_2Se_3$ as given in Fig.(\ref{e-ph_parameters}).

\subsection{A-In-plane spin texture anomaly}
\begin{figure}[h]
% produced by the gplt_fig_delta-E_Batanouny2 gnuplot file and the SC_sol_Batanouny_pot_LT_12 data file 
\includegraphics[scale=0.38]{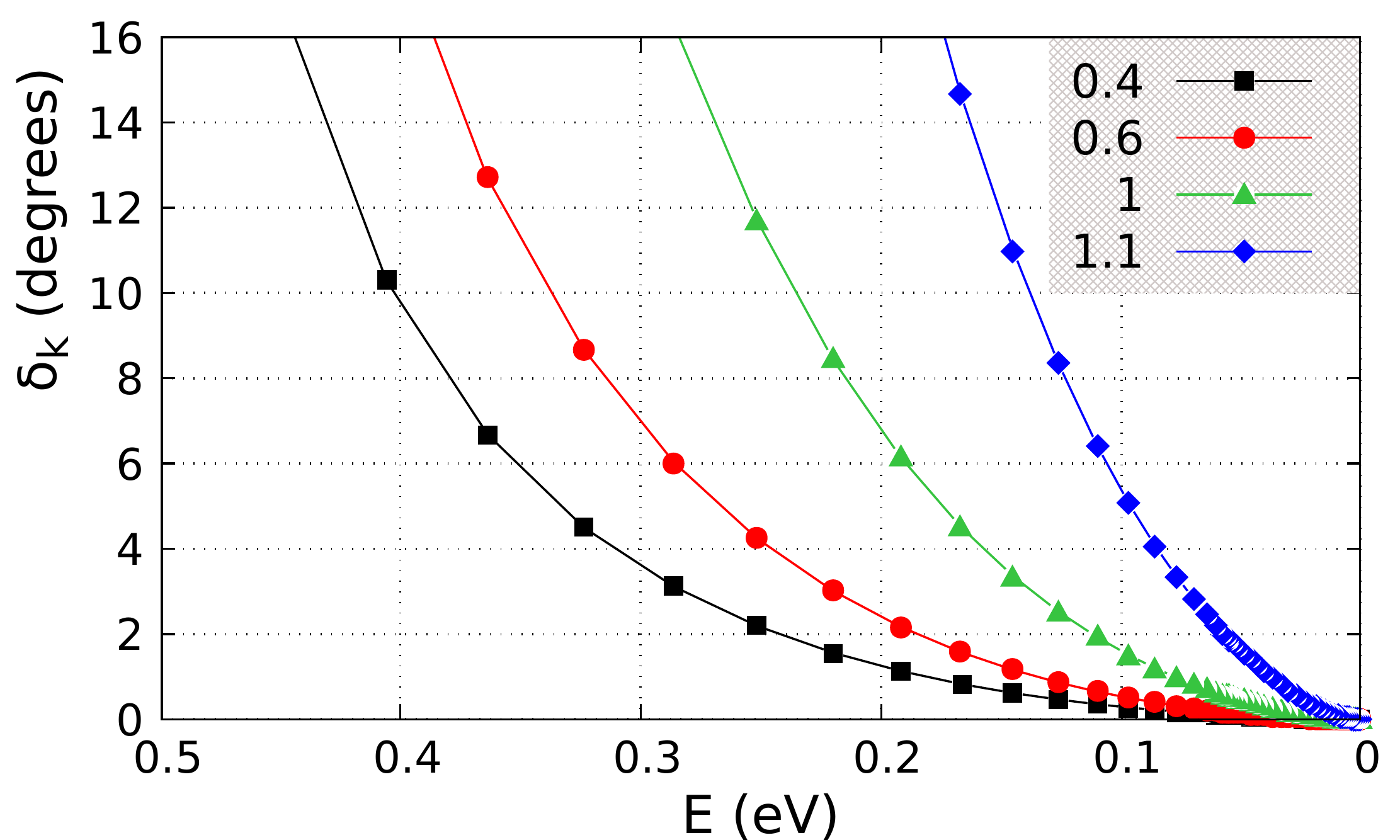}
\vskip-0.3truecm
\centering
\caption{ The inplane spin deviation angle $\delta_{\bs k}$ along the $\bar{\Gamma}\bar{M}$ direction as a function of energy due to the e-ph interaction in Ref.\cite{Batanouny2}. The legend indicates the values of $f_0$.}
\label{delta_E_Batanouny_data}
%\vskip-0.3truecm
\end{figure}

With the interaction parameters determined, we now introduce the anomalous in-plane spin deviation angle $\delta_{\bs k}$. This is a measure of the deviation in the surface spin-texture from the orthogonal spin-momentum locked configuration defined by 
\be
\sin \delta_{\bs k}=\frac{{\bs S}_{\bs k}.{\hat{\bs k}}}{\vert {\bs S}_{\bs k}\vert}=\frac{T_{I\bs k}}{\vert {\mathfrak g}_{\bf k}\vert }
\label{sin_delta_k}
\ee
The second equality follows from Eq.(\ref{spin_1.6}). The numerical results for $\delta_{\bs k}$ as a consequence of Eq's(\ref{polar_notation_1}-\ref{polar_notation_3}) are shown in Fig.(\ref{delta_E_Batanouny_data}). 
The devitation only extends to a few degrees in the Dirac regime $0 \le E (eV)\le 0.1$. This is followed by a region of sharp monotonic increase until the Fermi energy. As a result of Fig.(\ref{delta_E_Batanouny_data}) and the experimental measurement of $\delta_{\bs k}$ for $Bi_2Se_3$ in Ref.\cite{Y.H.Wang} we conclude that, the optical EPI reported in Ref.\cite{Batanouny3,Batanouny4} brings an explanation of the in-plane STA. The deviation angle reaches as high as $20$ degrees close to the $E_F\simeq 0.3 eV$. At this point, a strong hybridization with the bulk states is known to exist and additional phonon branches can be effective in the experimental results\cite{Sobota}. 

In Ref.\cite{Y.H.Wang} an unexpected rise in $\delta_{\bs k}$ is also reported in the Dirac regime. This  surprising finding is unlikely to be real effect because this region is energetically well isolated from the bulk electronic bands. A plausible explanation was already provided in Ref.\cite{Y.H.Wang} that, a small phase space area and the singularity in the spin-1/2 vortex at the tip of the Dirac cone combined with the limited experimental resolution can bring large inaccuracy in the measurement of the spin direction. 

Finally, the in-plane STA is a consequence of a cooperation between the interactions, strong SOC and the warping anisotropy. In this respect, the anomaly is not specific to this compound but a general consequence of the optical EPI in the same family of compounds. 

There are other experimental puzzles in this family related to the out-of-plane spin polarization. It is interesting to examine whether the interactions can also shed some light for their qualitative understanding which we now study.      

\subsection{B-Interaction effects in the out-of-plane spin}
Recent experiments made by the TIs $Pb(Bi,Sb)_2Te_4, Bi_2Te_3,Bi_2Se_3$ and $TlBiSe_2$ reported new puzzling results about the out-of-plane spin polarization\cite{Nomura}. In these experiments the spin polarization $P_z$ of the surface state was measured and normalized with the total polarization $P$ in order to minimize the bulk contamination.  Theoretically, $P_z/P$ and $S_{z}({\bs k})$ are related by  
\be
S_{z}({\bs k}) \longrightarrow \frac{1}{2} \,\frac{P_z}{P}~.
\label{out_of_plane_spin_polarization}
\ee
Two important conclusions were drawn in these measurements. The first is that, the out-of-plane spin is found to be correlated with the HW strength. Secondly, in some materials like $Bi_2Se_3$, the $P_z/P$ is reported to agree with the predictions of the IPP, whereas in some others, like $Bi_2Te_3, Pb(124)$ and $Pb(147)$, it can stay less than unity even in large momenta. This is not well understood because, according to the predictions of the k.p theory, the $P_z/P$ is expected to saturate at unity in large momenta due to the dominant role of the HW. 

In Ref.\cite{Nomura} a number of possibilities were examined as likely origin of this puzzle among which are the spin-dependent scattering\cite{Jozwiak} and the uncertainty between the phenomenological values of the paremeters of the energy spectrum. The former possibility is unlikely since they used unpolarized light in their experiment. Secondly, the uncertainty in the spectral parameters was questioned from the perspective of the neglected interaction effects in the theory. It is already known that the interactions do not lead to appreciable renormalization of the Dirac spectrum other than the finite quasiparticle lifetime effect and our observation in Section.III is also in agreement with this result. With all other possibilities eliminated, the authors of Ref.\cite{Nomura} left room for an explanation based on interaction effects. Based on our intuition acquired in Section IV-A, we consider that, the explanation of the $P_z/P$ anomaly may rest in the SDSE sector studied in this article.

In the light of these facts, we are prompted to revisit the out-of-plane spin in the context of this article. Our aim is to focus on the anomalous less-than-unity value of the $P_z/P$ in large momenta. For this, the relevant quantity in our context here is $\Sigma^*_{z\bs k}$. From Eq.(\ref{2D_sigma_z_T_k_1_a}), the leading term is 
\be
\Sigma^*_{z\bs k} \simeq \mathscr{C}_k \cos 3\phi 
\label{out-of-plane_1}
\ee
where $\mathscr{C}_k$ is a radial function and an isotropic term is forbidden by symmetry. The higher order terms can be present but neglected here. The $\mathscr{C}_k$ is found as a result of the Eq's.(\ref{polar_notation_1}-\ref{polar_notation_3}) and the resulting $\Sigma^*_{z\bs k}$ is plotted in Fig.(\ref{SE_xyz_solution}). This can be used in the calculation of the $S_z({\bs k})$ hence a theoretical prediction for $P_z/P$ can be made using Eq.(\ref{spin_1.3}) and (\ref{out_of_plane_spin_polarization}) as,         
%\be
%$S_{z\bs k}=(\hbar/2)\,\hat{G}^*_z({\bs k})$
%\label{out-of-plane_spin_1}
%\ee  
\be
\frac{P_z}{P}=\frac{g_{z\bs k}+\Sigma^*_{z\bs k}}{\vert {\bs G}^*({\bs k})\vert }\,.
\label{out_of_plane_2}
\ee
The deviation in $P_z/P$ from unity in large momenta is determined by the competition between the in-plane and the out-of-plane components of the SE in $\vert {\bs G}^*({\bs k})\vert=[(g_{xy\bs k}+\Sigma^*_{xy\bs k})^2+(g_{z\bs k}+\Sigma^*_{z \bs k})^2]^{1/2}$. 
\begin{figure}[h]
%produced by out_of_plane_HD_attr_C6 and plotted by out_of_plane_gplt_fancy_dene
\includegraphics[scale=0.39]{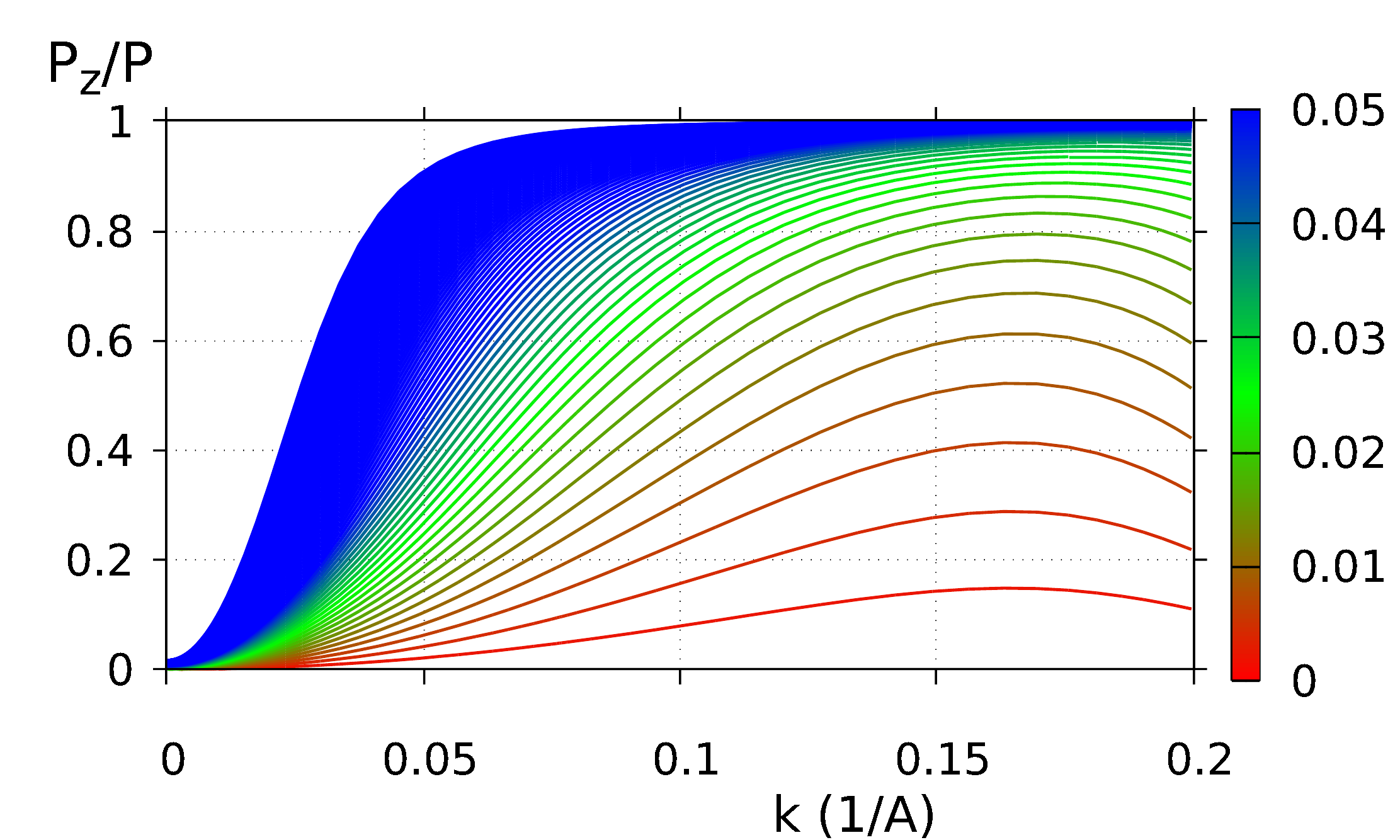} %out_of_plane_1.pdf}    
\vskip-0.3truecm
\caption{The out-of-plane spin polarization in Eq.(\ref{out_of_plane_2}) for fixed HW and SOC constants and various interaction strengths. The color scale shown on the right describes the dimensionless HW constant ${\bar \lambda}_0=\lambda_0/(E_D d^3)$.}  
\label{out_of_plane_spin_2}
%\vskip-0.3truecm
\end{figure}
We find in Fig.(\ref{out_of_plane_spin_2}) that, the $P_z/P$ ratio is very sensitively dependent on the relative strengths of the HW and the interactions. If the $g_{z\bs k}+\Sigma^*_{z\bs k}$ is the leading term for large momenta, Eq.(\ref{out_of_plane_2}) is in agreement with the expectations of the IPP even in the presence of interactions. This occurs in the context here, when the overall energy scale of the HW is much larger than the interactions [depicted by the blue lines in the Fig(\ref{out_of_plane_spin_2})]. The weakness of the interactions in $Bi_2Se_3$ can be sufficient to satisfy this condition. On the other hand, the scenario is different if the overall HW strength is compatible or weaker than the interaction (from green to red region in the figure). The HW in $Te$ based material is much larger than the $Bi_2Se_3$, but the overall interaction effects are reported to be much larger than in the $Se$ based compound. This can lead to the agreement with the IPP in the $Se$ case, whereas deviations in the $Te$ based compound from the IPP predictions. 

The Fig.(\ref{out_of_plane_spin_2}) provides only a qualitative understanding of the sub-unity values of $P_z/P$ in terms of interactions. A thorough experimental study on the mode specific interactions is needed for the whole $Bi_{2-y} Sb_y Se_x Te_{3-x}$ family before a satisfactory quantitative understanting is reached.  
  
\section{V- Conclusion}
Currently, "interaction" is a word of caution in topological materials due to the overwhelming success of the independent particle dynamics in its predictions and understanding the band structure. On the other hand, the presence of interactions has been undeniably shown in a large number of experiments only some of which have been pointed out in this work. The common feature of the topological materials is the presence of a strong spin-orbit coupling and this renders the spin as an invaluable source for completely understanding these materials. It is also not accidental that a large number of challenges to the IPP are connected with the spin related phenomena. In this article we demonstrated that the anomalies in the spin direction can be considered as signatures of interactions and the measurements of the spin-texture can be used for probing the interactions.    

Similar anomalies in the in-plane and out-of-plane spin data are being reported not only in TIs but also in other Dirac materials. For instance, the anomalous component of the spin parallel to the in-plane momentum was very recently observed in $\alpha$-$Sn$ which has a different point group symmetry than the TI family studied in the current work. This material is reported to be a Dirac semimetal or a strong TI depending on the internal strain\cite{Scholz,tetragonal_Dirac_semimetals}. It is an exciting challenge if the theory presented in this work may be relevant for a larger class of Dirac materials with strong SOC\cite{Dirac_materials}.

\begin{acknowledgements}
The support of the Boston University, Department of Physics is acknowledged where the majority of this work was done. The author is grateful to M. El-Batanouny for providing the experimental data. He has special thanks to A. Polkovnikov, M. El-Batanouny and A. Bansil for useful discussions. This work would not have been possible without the numerical parallelization of the Eq's(\ref{polar_notation_1})-(\ref{polar_notation_3}) accomplished by B. Karaosmanoglu and O. Ergul in METU, Department of Electrical Engineering. 
\end{acknowledgements}

\begin{appendix}
\section{Appendix}
We derive Eq.(\ref{spin_1.3}) in this section. Using the Dyson equation in (\ref{Dyson_eq}) and (\ref{Self-energy}),  
\be
\underline{\cal G}({\bs K})=\bigl[(i\nu_m-\xi_{\bs k}) - {\bs G}({\bs K}).{\bs \sigma}\bigr]^{-1}~,
\label{app1.1}
\ee 
where ${\bs G}({\bs K})=[{\mathfrak g}_{\bs k}+{\bs \Sigma}({\bs K})]$ and we ignored the SNSE correction. Converting the $\underline{\cal G}({\bs K})$ to the retarded GF with $i\nu_m \to \nu_-=\nu-i\delta$ as $\underline{\cal G}({\bs k},\nu_-)$ the Eq.(\ref{spin1.2}) becomes  
\be
{\bs S}({\bs k})=\frac{1}{2}\lim_{\delta \to 0^+}\int \limits_{-\infty}^\infty \frac{d\nu}{2\pi}{\it Tr} \Bigl\{\underline{\cal G}({\bs k},\nu_-).{\bs \sigma}\Bigr\}
\label{app1.2}
\ee
After some algebra with the ${\bs \sigma}$ matrices, the Eq.(\ref{app1.2}) becomes, 
\be
{\bs S}({\bs k})=\frac{1}{2}\lim_{\delta \to 0^+}\int \limits_{-\infty}^\infty \frac{d\nu}{2\pi}D(\nu_-) {\it Tr}\Bigl\{
\nu_--\xi_{\bs k}+{\bs G}_-.{\bs \sigma}\Bigr\}~~~~~~
\label{app1.2_1}
\ee
where ${\bs K}_-=({\bs k},\nu_-)$ and ${\bs G}_-={\bs G}({\bs K}_-)$. Here,
\be
D(\nu_-)=\frac{1}{2\vert {\bs G}_-\vert}\Biggl[ \frac{1}{\nu_--\xi_{\bs k}-\vert {\bs G}_-\vert}-\frac{1}{\nu_--\xi_{\bs k}+\vert {\bs G}_-\vert}\Biggr]~. \nonumber \\
\label{app1.4}
\ee
At this point we assume that the self energy  vector ${\bs\Sigma}({\bs K})$ has no new singularities or branch-cuts and its effect is only to shift the pole positions in Eq.(\ref{app1.4}) from the original positions $\xi_{\bs k}\pm \vert {\mathfrak g}_{\bs k}\vert $ in the absence of interactions. Inserting 
(\ref{app1.4}) in Eq.(\ref{app1.2_1}) and performing the trace at this pole position, we obtain 
\be
{\bs S}({\bs k})=\frac{1}{2}\hat{\bs G}^*({\bs k})
\label{app1.5}
\ee
where $\hat{\bs G}^*({\bs k})=\hat{\bs G}({\bs k})\Big\vert_{\nu=\nu_{\bs k}^*}$. The result obtained in Ref.(\cite{TH_STA}) is an approximate version of Eq.(\ref{app1.5}) valid within the Hartree-Fock scheme when $\underline{\Sigma}({\bs k})$ is independent of frequency. 
\end{appendix}

\end{document}